\documentclass{article}
\usepackage{spconf,amsmath,graphicx,url,tipa,booktabs,multirow}

\usepackage[font=small]{caption}  

\title{Unsupervised Accent Adaptation Through Masked Language Model Correction of Discrete Self-Supervised Speech Units}

\name{Jakob Poncelet, Hugo Van hamme
      \thanks{Research supported by Research Foundation Flanders (FWO) under grant S004923N of the SBO programme.}}
      
\address{KU Leuven \\ 
     Department Electrical Engineering ESAT-PSI, Leuven, Belgium\\
     }

\begin{document}
\ninept  

\maketitle

\begin{abstract}  
Self-supervised pre-trained speech models have strongly improved speech recognition, yet they are still sensitive to domain shifts and accented or atypical speech. Many of these models rely on quantisation or clustering to learn discrete acoustic units. We propose to correct the discovered discrete units for accented speech back to a standard pronunciation in an unsupervised manner. A masked language model is trained on discrete units from a standard accent and iteratively corrects an accented token sequence by masking unexpected cluster sequences and predicting their common variant. Small accent adapter blocks are inserted in the pre-trained model and fine-tuned by predicting the corrected clusters, which leads to an increased robustness of the pre-trained model towards a target accent, and this without supervision. We are able to improve a state-of-the-art HuBERT Large model on a downstream accented speech recognition task by altering the training regime with the proposed method.
\end{abstract}


\section{Introduction}
\label{sec:intro}
Recently, acoustic models pre-trained in a self-supervised fashion have made big waves in the field of speech processing. Hidden representations learned by HuBERT \cite{hsu2021hubert}, wav2vec 2.0 \cite{baevski2020wav2vec}, WavLM \cite{wavlm} and others \cite{pmlr-v162-chiu22a, Chung2021w2vBERTCC} can be used to achieve strong performance on a variety of speech processing tasks, without requiring copious amounts of task-specific labeled data. However, prior research has revealed that while these pre-trained acoustic models are able to discover informative speech units \cite{ma2021probing}, they are sensitive to domain shifts between the (pre-)training and target domain \cite{hsu2021robust}, to speaker variations \cite{qian2022contentvec} and to contextual factors like co-articulation and allophony \cite{hallap2022evaluating}. 

The learned pre-trained representations can be biased towards standardised canonical data characteristics of the pre-training corpora, and consequently perform worse on nonnative or accented speakers, as shown for HuBERT in \cite{Bhatia2023}. Following \cite{wells_1982}, dialectal changes can to some extent be described as a collection of vocal shifts from a standard pronunciation paradigm. As the learning process of many self-supervised pre-trained acoustic models crucially involves quantisation or clustering of latent features in the prospect of discovering informative speech units, this shift will also be represented in the discretised features.

In this paper, we propose to improve accent-invariance by learning to remove the acoustic accent shift from the pre-trained model representations, without any supervision. We model the standard pronunciation statistics by training a masked language model \cite{devlin-etal-2019-bert} on discrete speech units from the source language, generated by clustering the pre-trained model's hidden features. The extracted pre-trained discrete clusters of accented speech are then corrected towards the standard accent by applying an iterative mask-and-decode algorithm to the language model. Finally, we fine-tune the pre-trained models with these corrected clusters as self-supervised targets and show improved invariance towards the target accent.

Our approach requires only unlabeled, non-parallel data from the source language and the target accent. In this work, we consider North American English to be the standard source accent, as it is the conventional language in most work on English speech recognition (e.g. LibriSpeech \cite{librispeech}). We evaluate the approach for several English accents from the CommonVoice dataset \cite{Ardila2019CommonVA} on an automatic speech recognition (ASR) task. We perform continuous ASR with fine-tuned HuBERT model features and show improved Word Error Rates (WER) compared to the uncorrected models on accented speech.

\begin{table*}[t]
    \centering
    \resizebox{\textwidth}{!}{%
    \begin{tabular}{r | c c c c c c c c c c c c c c c c c c}
        \toprule
        Sentence & \multicolumn{18}{c}{WHERE IS ...} \\
        \midrule
        Input phones* & W & W & AH & AH & AH & AH & R & R & R & R & IH & IH & IH & Z & Z & Z & Z & Z \\
        Input clusters & 7 & 345 & 181 & 181 & 181 & 181 & 468 & 406 & 406 & 467 & 356 & 356 & 356 & 281 & 281 & 453 & 9 & 9 \\
        Masked inputs & 7 & 345 & [M] & [M] & [M] & [M] & 468 & 406 & 406 & 467 & [M] & [M] & [M] & 281 & 281 & 453 & 9 & 9 \\
        Output clusters & 7 & 345 & 109 & 109 & 264 & 264 & 468 & 406 & 406 & 467 & 356 & 356 & 356 & 281 & 281 & 453 & 9 & 9 \\
        Output phones* & W & W & W & W & EH & EH & R & R & R & R & IH & IH & IH & Z & Z & Z & Z & Z \\
        \bottomrule
    \end{tabular}
    }
    \caption{Example of the inputs and outputs of the MLM for part of an accented utterance in CommonVoice. Note that the input and output phone labels (*) are not used by the model, but generated afterwards as the most likely cluster to phone mapping. [M] denotes a mask.}
    \label{tab:phoneexample}
\end{table*}

\section{Preliminaries}  
\label{sec:method}

\subsection{BERT}
BERT \cite{devlin-etal-2019-bert} is a language model based on Transformers \cite{NIPS2017_3f5ee243} that learns strong bidirectional language representations. The main objective is masked language modeling: part of the input tokens are masked, and the model is optimised to fill in those masked tokens conditioning on both left and right context with attention layers. The masks can be applied randomly or in larger spans comprising several consecutive tokens, as in SpanBERT \cite{joshi-etal-2020-spanbert}. The output layer maps the feature representations back to the vocabulary of tokens. In the standard BERT model, masked language modeling is combined with a contrastive next sentence prediction objective. Finally, distilBERT \cite{sanh2019distilbert} is a smaller, distilled version of BERT. 

\subsection{HuBERT}
As audio and speech are continuous signals, a language model cannot be applied directly without a discretisation step. HuBERT \cite{hsu2021hubert}, or Hidden Unit BERT, is a popular approach for self-supervised speech representation learning from raw audio. First, an offline clustering step is applied to discover discrete acoustic units in the audio. Applying the K-means algorithm to MFCC features for example gives meaningful units. Second, the model is trained to predict these aligned discrete units from the partly masked output of an acoustic feature encoder using a bidirectional BERT-like encoder. Hence, HuBERT intrinsically combines an acoustic and masked language model over continuous features of units. Finally, the model is improved iteratively by re-training with better discrete units, obtained from clustering the encoder features of a previous iteration model. Note that a HuBERT Large model (24 layers) is pre-trained using the clusters from a 2nd iteration HuBERT Base model (12 layers) \cite{hsu2021hubert}.

\subsection{Adapters}
Fine-tuning a large pre-trained model on a new dataset or task can be very computationally expensive, as the scale of pre-trained models is increasing rapidly. Instead of fine-tuning the entire model, nearly the same improvements can be accomplished by only fine-tuning small residual blocks (i.e. adapters) that are added to every layer \cite{Houlsby2019ParameterEfficientTL} and freezing all pre-trained weights, drastically reducing the number of parameters to optimise. Furthermore, adapters alleviate catastrophic forgetting of learned knowledge during pre-training \cite{steven}, as the pre-trained weights are not altered.

\section{Approach}
This work aims to improve the accent-invariance of pre-trained models using unlabeled speech from the source and target accent. A masked language model (MLM) is trained on discrete speech units from the source accent, of which generally lots of data is available, and afterwards the model is used to correct the discrete units for an accented utterance.

\begin{figure}[htb]
    \centering
    \includegraphics[width=\columnwidth]{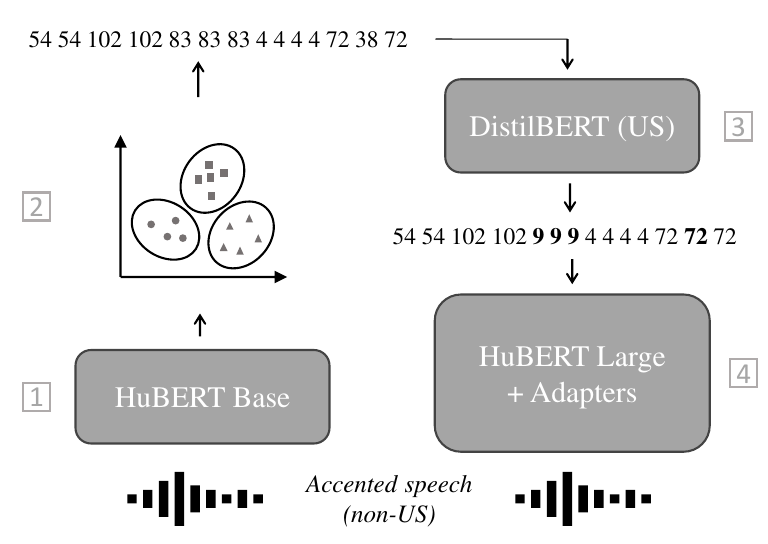}  
    \caption{Schematic of the proposed method. 1. Features are extracted for an accented utterance with the HuBERT Base model. 2. K-means quantisation to a discrete token sequence. 3. Clusters are corrected by the distilBERT model, trained on US data, using the mask and predict algorithm. 4. HuBERT Large model is adapted to the accent by optimising small adapter blocks during continual pre-training with the corrected clusters.}
    \label{fig:model}
\end{figure}

\subsection{Standardised Unit Language Model}
Offline clustering of the hidden features of a self-supervised pre-trained model leads to informative discrete speech units that correlate well with actual phonemes as in HuBERT \cite{hsu2021hubert}, without having seen any labels. Previous work has shown that these discrete acoustic units are strong building blocks for modeling spoken language without supervision \cite{lakhotia-etal-2021-generative, wav2vec-u}. To model the standard pronunciation in the source language, North American English, we train an MLM on the discretised utterances from that accent. To aid training, the MLM is initialised from a pre-trained distilBERT model\footnote{\url{https://huggingface.co/distilbert-base-uncased}}. The token embeddings are re-initialised as the vocabulary now only consists of 500 clusters id's. The MLM is optimised with the masked prediction objective, omitting the next sentence prediction. Following SpanBERT \cite{joshi-etal-2020-spanbert}, the masks are applied in larger spans masking several consecutive frames, as one frame corresponds to only 20ms of audio.

\subsection{Accented Cluster Correction}
Accents are characterised by a vocal shift from a standard pronunciation \cite{wells_1982}. Vowel shifts are the most obvious examples, but not the only ones. For example, notice the difference between British and North American pronunciations for the words 
\textit{dance} (\textipa{/dA:ns/} vs \textipa{/d\ae{}ns/}), \textit{stupid} (\textipa{/"stju:.pId/} vs \textipa{/"stu:.pId/}) and \textit{either} (\textipa{/"aI.D@\super{r}/} vs \textipa{/"i:.D\textrhookschwa/})\footnote{According to Cambridge Dictionary using IPA symbols.}. 

An acoustic model trained in a self-supervised way will identify these sounds as belonging to separate units, although they represent the same characters. To achieve accent-invariance, the masked language model trained on the standard pronunciation should correct this accent shift by changing the clusters that are affected back to the (statistically most-likely) standard pronunciation clusters, e.g. the \textit{a} in the above example \textit{dance}, taking context into account.

To correct a sequence of cluster id’s corresponding to an accented utterance with the MLM, we apply an iterative mask-and-decode algorithm, similar to Mask-Predict \cite{ghazvininejad-etal-2019-mask}. As the MLM is not conditioned on the audio, conditional decoding algorithms like Mask-Predict \cite{ghazvininejad-etal-2019-mask}, Mask-CTC \cite{Higuchi2020} and BERT-CTC \cite{higuchi-etal-2022-bert} cannot be trivially applied, and only a small fraction of the input sequence should be masked, or the model will not be able to recover the original sentence. Moreover, the masks are applied to groups of identical consecutive clusters, to correct entire speech units more easily.

The cluster rescoring algorithm works as follows. First, the input cluster sequence is forwarded through the MLM, and the frame-wise softmax output scores are used as confidences for the input tokens. Ideally, the aforementioned accent shifts are given a lower confidence by the LM, as it is trained to predict an alternative pronunciation for those sounds in a word. After scoring, consecutive identical clusters are grouped together. The score of every cluster group is determined by the maximum score of all clusters in the group, as in \cite{Higuchi2020}. Then, the cluster groups with the lowest scores are masked, so that at least $N_k$ frames are masked. Next, the masked inputs are digested by the MLM which gives it's predictions for the mask tokens. Finally, the $M$ most confident predictions of these cluster groups are filled in. The masks are computed again for the corrected clusters, and the decoding continues. This procedure is carried out for $K$ iterations, where the total number of masks adaptively shrinks every iteration $k$ as $N_k = \lfloor N_{max} \cdot \frac{K-k+1}{K}\rfloor$ \cite{Higuchi2020}. The number of masks $M$ that are filled in every iteration is fixed as $M = \frac{N_{max}}{K}$. The maximum number of masks is determined by the length $T$ of the utterance and the masking ratio $P_{mask}$, as $N_{max} = P_{mask} \cdot T$.

The number of iterations $K$ and the masking percentage $P_{mask}$ influence how much of the input sequence will be altered.

\subsection{Accent-Invariant Model Adaptation}

Previous work \cite{Bhatia2023} has shown that pre-trained HuBERT models benefit from adaptation to a new accent by continually pre-training on data from the target accent. To this end, appending accent-specific adapters to the encoder layers and fine-tuning only the adapter parameters, reaches almost the performance of full encoder fine-tuning but with a strongly reduced number of parameters.

This work builds upon \cite{Bhatia2023}, continually pre-training HuBERT on corrected cluster sequences by fine-tuning accent-specific adapters, improving accent-invariance without supervision. We apply Houlsby bottleneck adapters \cite{Houlsby2019ParameterEfficientTL} after every attention and feedforward block in the encoder. The adapters will learn to normalise the accent shifts that have been corrected by the MLM.

\section{Experiments}

\begin{table}[t]
    \centering
    \begin{tabular}{l c c}
        \toprule
        & \multicolumn{2}{c}{\textbf{PER}} \\
        \textbf{Masking strategy} & \textit{dev} & \textit{test} \\
        \cmidrule(lr){1-1} \cmidrule(lr){2-2} \cmidrule(lr){3-3}
        Original clusters & 85.69 & 85.76 \\
        Adaptive masking cluster groups & 80.55 & 80.61 \\
        Adaptive masking phone groups* & 76.09 & 76.14 \\
        Adaptive masking phone groups fill-all* & 63.92 & 63.95 \\
        \bottomrule        
    \end{tabular}
    \caption{Phone Error Rates of rescored British English utterances, converted from clusters to phones, w.r.t. a North-American English G2P transcription. For phone groups (*), masks are applied to groups of frames based on their phone class. \textit{Fill-all} means that all masks are filled in every iteration, instead of a fraction.}
    \label{tab:phones}
\end{table}

\begin{table*}[]
    \centering
    \begin{tabular}{l c c c c c c c c }
    \toprule
     & \multicolumn{4}{c}{ASR w/ LS960} & \multicolumn{4}{c}{ASR w/ CV-US} \\
    \cmidrule(lr){2-5} \cmidrule(lr){6-9}
    \textbf{Model / Masking strategy} & \textit{dev-clean} & \textit{test-clean} & \textit{dev-eng} & \textit{test-eng}  & \textit{dev-clean} & \textit{test-clean} & \textit{dev-eng} & \textit{test-eng} \\
    \cmidrule(lr){1-1} \cmidrule(lr){2-2} \cmidrule(lr){3-3} \cmidrule(lr){4-4} \cmidrule(lr){5-5} \cmidrule(lr){6-6} \cmidrule(lr){7-7} \cmidrule(lr){8-8} \cmidrule(lr){9-9}
    HuBERT Large \cite{hsu2021hubert} & \textbf{5.77} & \textbf{6.01} & 24.39 & 19.82 & \textbf{6.84} & \textbf{7.22} & 16.79 & 14.26 \\
    HuBERT + Adapters \cite{Bhatia2023} & 6.14 & 6.50 & 22.11 & 17.92 & 7.18 & 7.64 & 15.87 & 12.45 \\  
    Adaptive masking cluster groups & 6.18 & 6.72 & 21.52 & 18.00 & 7.49 & 7.85 & 15.41 & \textbf{11.71} \\
    Adaptive masking only vowels if $k>3$ & 6.17 & 6.58 & \textbf{21.01} & \textbf{17.47} & 7.34 & 7.63 & \textbf{15.06} & 11.89 \\
    \bottomrule
    \end{tabular}
    \caption{ASR experiments with WER results. The ASR model is either trained on LibriSpeech (LS960) or on the North-American English split of CommonVoice (CV-US), and evaluated on LibriSpeech dev-clean/test-clean or on the British English (ENG) part of the CommonVoice dev/test set. Row 3 indicates a model fine-tuned with the outlined algorithm (K=10). Row 4 first applies the masking algorithm to make basic corrections in the first 3 iterations, and focuses on vowel shifts by only masking vowels in the following iterations (based on the cluster id).}
    \label{tab:asr_eng}
\end{table*}

\begin{table*}[t]
    \centering
    \resizebox{\textwidth}{!}{%
    \begin{tabular}{l c c c c c c c c c c c c}
        \toprule
         & \multicolumn{2}{c}{\textbf{US}} & \multicolumn{2}{c}{\textbf{ENG}} & \multicolumn{2}{c}{\textbf{AUS}} & \multicolumn{2}{c}{\textbf{IND}} & \multicolumn{2}{c}{\textbf{IRE}} & \multicolumn{2}{c}{\textbf{SCO}} \\
        \cmidrule(lr){2-3} \cmidrule(lr){4-5} \cmidrule(lr){6-7} \cmidrule(lr){8-9} \cmidrule(lr){10-11} \cmidrule(lr){12-13}
         \textbf{Model} & \textit{dev} & \textit{test} & \textit{dev} & \textit{test} & \textit{dev} & \textit{test} & \textit{dev} & \textit{test} & \textit{dev} & \textit{test} & \textit{dev} & \textit{test} \\
        \cmidrule(lr){1-1} \cmidrule(lr){2-2} \cmidrule(lr){3-3} \cmidrule(lr){4-4} \cmidrule(lr){5-5} \cmidrule(lr){6-6} \cmidrule(lr){7-7} \cmidrule(lr){8-8} \cmidrule(lr){9-9} \cmidrule(lr){10-10} \cmidrule(lr){11-11} \cmidrule(lr){12-12} \cmidrule(lr){13-13} 
        HuBERT Large \cite{hsu2021hubert} & 17.29 & 18.41 & 16.79 & 14.26 & 10.34 & 13.63 & 24.96 & 25.84 & 14.02 & 10.56 & 15.09 & 16.12 \\
        HuBERT + Adapters \cite{Bhatia2023} & 15.09 & \textbf{15.82} & 15.87 & 12.45 & \textbf{8.93} & 11.97 & 17.87 & 18.04 & \textbf{12.20} & 10.54 & 11.67 & 12.93 \\
        Ours MLM K=5 & 15.09 & 16.04 & 15.48 & \textbf{11.14} & 9.11 & 11.72 & 18.18 & 18.90 & 12.24 & 10.63 & \textbf{10.99} & 12.53 \\
        Ours MLM K=10 & \textbf{15.05} & 16.05 & \textbf{15.41} & 11.71 & 8.98 & \textbf{11.66} & \textbf{17.44} & \textbf{17.97} & 12.31 & \textbf{10.45} & 11.08 & \textbf{12.35} \\
        \bottomrule
    \end{tabular}
    }
    \caption{ASR experiments with WER results for multiple English accents. The proposed models are adapted to the target accent in every column separately. The ASR model is trained on the North-American English (US) split of CommonVoice, and evaluated on the target accent. The accents are abbreviated as US (North-American), ENG (British), AUS (Australian), IND (Indian) and SCO (Scottish).}
    \label{tab:asr}
\end{table*}

\subsection{Experimental Setup}
We use a subset of the CommonVoice \cite{Ardila2019CommonVA} 7.0 release with 6 well represented English accents for pre-training. The training set consists of 827k utterances (611 hours) of English speech, including North American (350k), British (116k), Indian (71k), Australian (56k), Scottish (11k) and Irish (7k) accents. There are 1k utterances from each accent for validation and testing. 

The baseline HuBERT Large model consists of 7 convolutional feature extraction layers and 24 Transformer layers, with a total of 316M parameters. It has been pre-trained on the 60k hour LibriVox corpus with target clusters extracted from the 9th layer of the 2nd iteration HuBERT Base model (12 Transformer layers), which was pre-trained on 960 hours of LibriSpeech. The K-means quantiser that generated the clusters has 500 centroids learned on LibriSpeech train-clean-100h. Both models and the quantiser are officially released checkpoints in fairseq\footnote{\url{https://github.com/facebookresearch/fairseq/tree/main/examples/hubert}}. 

For continually pre-training the HuBERT Large model, we insert Houlsby adapters at every Transformer layer, after the attention block and after the feed-forward block. The adapters consist of a feedforward down-projection layer, a ReLU, a feedforward up-projection layer and a layer normalisation block. The bottleneck dimension is set to 1024, which performed best in \cite{Bhatia2023}. The target clusters are extracted with the same HuBERT Base 2nd iteration model and the respective quantiser. The learning rate is set to 1.5e-3 and the model is updated for 30k steps with a linear learning rate schedule with 5k warm-up steps. We use a batch size of 32 with a maximum of 300k tokens. For a dataset of 100h, this setup corresponds to 30 full epochs. The adapter-HuBERT model is fine-tuned on only one accent per experiment.

The distilBERT masked language model consists of 6 Transformer layers. It is trained on masked token sequences where spans of 10 tokens are masked and on average 20\% of the entire utterance is masked ($P_{mask}=0.2$). Following BERT, masked tokens have a 20\% chance to be replaced with either the original or a random token, to reduce the train-test mismatch. The MLM is trained for 18 epochs with a batch size of 32, or about 200k steps. The learning rate follows a linear decay starting from 5e-5. The MLM is trained on the North American English split of the dataset.

The ASR model consists of a 2-layer BiLSTM (as in the SUPERB challenge \cite{yang21c_interspeech}) with 512 hidden units, corresponding to around 10M parameters. It is optimised with a CTC loss \cite{graves_CTC} to predict characters. The input to the ASR model is a learnable weighted sum of the pre-trained model's Transformer layer hidden features. SpecAugment is applied to the inputs and the model is trained for 100k steps with a learning rate of 1e-4. The ASR model is trained on the North American English split of the dataset, and evaluated on all English accents to show unsupervised accent adaptation.


\subsection{Phone Modeling}
\label{sec:phon}
In this first experiment, we evaluate the cluster corrections of the MLM towards the North-American accent. As there are only 500 different clusters, the discrete units are better interpretable. Based on force-aligned phoneme annotations\footnote{\url{https://zenodo.org/record/2619474}} for LibriSpeech-100h, we determine the most likely mapping from the cluster id to the corresponding phoneme (40 classes). Then, we use a North-American G2P\footnote{\url{https://pypi.org/project/g2p-en/}} on the transcripts of the accented dev and test set to get a reference phoneme transcription for these utterances. We rescore the original clusters with the MLM and the proposed mask-decode algorithm to get corrected clusters. Next, we map both the original clusters and the rescored cluster sequences to their phoneme equivalent based on the learned cluster-phoneme mapping. Table \ref{tab:phoneexample} shows an example of input and output clusters and their mapping to phonemes. Finally, we evaluate the Phoneme Error Rate (PER) between the reference G2P phonemes and the original/corrected phonemes. Results are shown for the British English utterances in the dataset in Table \ref{tab:phones}, applying the masking algorithm either to cluster groups or to phone groups, where the phone group is derived from the cluster id based on the learned mapping.

\subsection{Accented Speech Recognition}
The main goal of this research is improving the invariance of a pre-trained model towards an accent without having labeled data for that accent. We train an ASR model on North-American English data and evaluate it on the other accents, using the features of the adapted HuBERT models for every accent separately. For the British English accent, Table \ref{tab:asr_eng} displays the results for different masking schemes, for an ASR model trained on the US part of CommonVoice (accent mismatch) or on LibriSpeech (accent and dataset mismatch). For one of the results, the MLM focuses on replacing vowel shifts, if there is some labeled source data available (as in \ref{sec:phon}). Table \ref{tab:asr} shows the WER results for the experiments on all accents with an ASR model trained on CommonVoice, where the outlined unsupervised mask-and-predict algorithm is applied to cluster groups for $K$ iterations.


\section{Discussion}
\textbf{Phone modeling} As the PER w.r.t. the US G2P transcription drops, the clusters are closer to the North American pronunciation. Empirically, we see the MLM corrects some vowels and deviant pronunciations, and reduces cluster artifacts by smoothing similar clusters into one group. However, changing too much can be harmful, and there is no real ground truth for cluster correctness. There is a trade-off between smoothing and removing insertions on the one hand, and keeping temporal detail for ASR purposes on the other hand.

\textbf{Speech recognition} The experiments show that for most accents there is an improvement over regular fine-tuning of HuBERT with uncorrected clusters of \cite{Bhatia2023}, indicating an improved accent invariance by rescoring the clusters. The idea of modeling the discrete units with a language model on canonical data and correct atypical speech is conceptually simple and applicable to many models. If there is labeled data from the source accent available, the clusters can be mapped to phones and accent-specific characteristics can be targeted (e.g. vowel shifts between British and American English).

\section{Related Work}
Several works have built on the idea of combining discretised self-supervised learning with a language model \cite{lakhotia-etal-2021-generative}. In vq-wav2vec \cite{BaevskiSA20, baevski2020effectiveness}, the learned quantiser of a pre-trained wav2vec model is used to generate discrete audio codes, which are then modelled by a BERT model. Afterwards, the continuous BERT features are used for downstream tasks like ASR. There are several differences with our work. First of all, they quantise into 12.5k unique codes, which makes the BERT training more difficult (128 GPU’s) and the clusters less interpretable. Furthermore, we use a novel mask-and-decode algorithm so that the rescored clusters can be used to adapt a HuBERT model or for discrete lightweight ASR \cite{chang2023exploration}. Finally, we make adaptations to specifically target the case of accented speech.

Most research on accent robustness has focused on adaptation in multi-accent speech recognition models \cite{Shi2021TheAE, lwfaeemas, multiaccent}. Other recent work \cite{zhao2022SLT} has investigated codebook regularisation to phonemes in wav2vec 2.0 to improve accent invariance during pre-training without supervision. Finally, \cite{aipnet} uses discriminators to disentangle accent-specific information during auto-encoder pre-training.

\section{Conclusion}
This work improves the accent-invariance of a state-of-the-art HuBERT Large model by correcting the discrete acoustic units towards standard pronunciation, with a simple adaptation of its training scheme. A masked language model is trained on the discrete units of a common accent, and corrects the accented token sequences with an iterative mask-and-decode algorithm. Fine-tuning the pre-trained model with adapters on these corrected clusters improves speech recognition for the target accent, without any labels.

In future work, we plan to explore combining multiple adapters to enable multi-accent training and be resilient against L2 speakers, and experiment with bigger language models. It is also interesting to combine speech and MLM encoders in an end-to-end fashion.

\vfill\pagebreak

\bibliographystyle{IEEEbib}
\bibliography{strings,refs} 

\end{document}